\documentclass[a4paper,11pt]{article}
\usepackage{graphicx}
\usepackage{hyperref}
\usepackage{amsmath}
\title{The impact of quantum computing on real-world security: A 5G case study}
\author{Chris J. Mitchell\\Information Security Group, Royal Holloway, University of London\\
\url{www.chrismitchell.net}}
\date{13th December 2019}
\begin{document}

\maketitle

\section*{Abstract}

This paper provides a detailed analysis of the impact of quantum computing on
the security of 5G mobile telecommunications. This involves considering how
cryptography is used in 5G, and how the security of the system would be
affected by the advent of quantum computing.  This leads naturally to the
specification of a series of simple, phased, recommended changes intended to
ensure that the security of 5G (as well as 3G and 4G) is not badly damaged if
and when large scale quantum computing becomes a practical reality.  By
exploiting backwards-compatibility features of the 5G security system design,
we are able to describe a multi-phase approach to upgrading security that
allows for a simple and smooth migration to a post-quantum-secure system.

\section{Introduction} \label{Intro}

In the last few years that has been much discussion of the
impact of quantum computing on cryptographic systems.  Whilst
there is no general agreement that large-scale, general
purpose, quantum computers will ever be built --- see, for
example, Dyakonov \cite{Dyakonov19} --- a huge effort continues
to be made to develop them.  Also, as has been widely
discussed, should such computers ever become available, the
impact on the security of many of today's cryptographic systems
is considerable.

This suggests that for every major application of cryptography a careful review
of the impact of quantum computing needs to be performed without delay. This
review should assess which parts of the system are vulnerable if a quantum
computer should become available, and what the impact would be if this part of
the system is broken. Such a review should also consider how long it would take
to replace the cryptography used in each part of the system; this assessment
should include the time required to update the specifications, the time to
produce replacement implementations, and the time to replace all the existing
implementations `in the field'. The total time could be very considerable,
depending on the application domain.  For example, credit and debit cards have
a typical lifetime of three--five years, so that replacing all such cards with
new technology could take a decade or more (and this doesn't even consider the
time required to replace the infrastructure supporting their use).

The security of the mobile telecommunications infrastructure has relied on the
use of cryptography since the advent of the GSM system, originally designed
back in the 1980s and first deployed in 1991 (GSM is often referred to as
\emph{2G} for the \emph{2nd generation} of mobile telecommunications). 5G is
the latest generation of mobile telecommunications standards, produced by the
3GPP organisation\footnote{\url{https://www.3gpp.org/}}, and 5G systems are now
being deployed globally.  Mobile telecommunications systems are very widely
used worldwide, and 5G looks set to become even more closely integrated with
society.  This means that the security of 5G systems is a matter of huge
importance.

These observations have motivated this paper, which addresses the impact of
quantum computing on 5G mobile security. Such a review seems particularly
timely, given that 5G technology is already being deployed and major
investments are likely to occur in this technology in the next few years. As
will become clear, key parts of the system as currently specified are
vulnerable should a quantum computer become available.  Performing this
detailed analysis enabled a phased approach to upgrading the security of the
current system to be proposed, which will allow a smooth and simplified
migration path.

Apart from the lessons we can draw from this review of priorities in 5G
security evolution, it is hoped that this study will also help in performing
`post-quantum' reviews for other widely deployed technologies reliant on
cryptography for security.

The remainder of this paper is organised as follows. \S\ref{quantum} provides
an introduction to the impact of quantum computing on the security of modern
cryptography.  This is followed, in \S\ref{5G}, by a (simplified) description
of how cryptography is used in 5G\@.  This provides the basis for the detailed
analysis in \S\ref{analysis} of the impact of quantum computing on 5G security.
This in turn leads to \S\ref{future} in which we provide a series of
prioritised recommendations regarding how 5G security should be developed to
ensure that it is not seriously affected by the advent of quantum computing.
Finally, a summary and conclusions are given in \S\ref{conclusions}.

\section{The impact of quantum computing on modern
cryptography}  \label{quantum}

\subsection{Cryptanalysis}  \label{grover}  \label{shor}

If and when large-scale general-purpose quantum computers are
constructed, the effect on currently used cryptography will be
very significant --- see, for example, Yanofsky and Mannucci
\cite{Yanofsky08}. In particular, two quantum algorithms (i.e.\
algorithms which will execute on a quantum computer) have been
devised which affect both symmetric (secret key) and asymmetric
(public key) cryptographic algorithms.

Shor's algorithm \cite{Shor94}, published in 1994, has a very major impact on
the security of all today's widely used asymmetric algorithms.  Complementing
this, Grover's 1997 algorithm \cite{Grover97} affects the security of any
symmetric algorithm, albeit in a much less severe way.  The impact of these
algorithms, given a quantum computer on which to execute them, can be
summarised as follows.
\begin{itemize}
\item All asymmetric cryptographic algorithms based on the difficulty of
    factoring large integers or computing discrete logarithms (include
    elliptic curve schemes) will be rendered insecure for currently used
    key lengths. As a result, all currently used asymmetric schemes could
    be broken using a large-scale general-purpose quantum computer.
    Moreover, the (large) increase in key length needed to make currently
    used schemes secure would be so great as to render use of the
    algorithms infeasible in most cases.
\item All symmetric cryptographic algorithms will in effect have their key
    length significantly reduced (in principle it is halved, but in
    practice the reduction will be somewhat less than this). That is, with
    the aid of a quantum computer, a $k$-bit key for a symmetric algorithm
    could be discovered in of the order of $2^{k/2}$ computations, i.e.\
    the square root of the order of $2^k$ computations required using a
    conventional computer.  However, in the quantum case these
    `computations' may be quite complex, which is why a simple square root
    argument is not quite correct --- also quantum attacks are not
    parallelisable in the way that brute force searches using conventional
    computers are. However, given the degree of uncertainty involved in
    estimating the precise quantum attack complexity, we take a
    conservative approach; that is, in line with established practice, we
    follow the principle that, if and when a quantum computer is available,
    a 128-bit key will offer roughly the same level of security as a 64-bit
    key does today, i.e.\ it will not be secure. Following this principle,
    to achieve the same level of security as provided today by a 128-bit
    key will require switching to 256-bit keys.
\end{itemize}
As a convenient shorthand, we refer to the putative point in
time when any potential adversary gains access to large-scale
general-purpose quantum computers as the \emph{PQ era}.

\subsection{Security impacts}

It is interesting to consider what the future impact will be if
cryptographic algorithms in current use are rendered insecure
at some time in the future. For algorithms solely used for
verifying the integrity of transmitted data, i.e.\ whose use
has no long-term impact, there will no significant problems, as
long as new `secure' algorithms are introduced in time.
However, the significance for encryption and key establishment
algorithms is potentially catastrophic.

If ciphertext is intercepted and stored that was encrypted using an algorithm
that becomes insecure at some future time, then it could be decrypted at the
future point in time when the encryption algorithm is broken.  That is, data
whose secrecy has any kind of long-term significance is being made vulnerable
right now through the use of algorithms which will become insecure if quantum
computers are built.

Similarly, suppose a key establishment technique is employed
which could be broken at some future time using data exchanged
via an insecure link.  If the key establishment exchanges are
recorded, then any sensitive data encrypted generated using a
key established using such a technique would be at risk of
compromise as and when a quantum computer becomes available.

\subsection{Replacement algorithms}  \label{replacement}

Given the potentially devastating consequences for the security of today's
cryptography, a number of agencies are working on developing new cryptography
standards. Fortunately, the current state of the art in symmetric cryptography
allows for use of 256-bit keys, so no major changes are needed in this domain,
except for enforcing a move to longer keys (i.e.\ of 256 bits or more).
However, the situation is not so simple for asymmetric cryptography, given that
almost all current applications of asymmetric cryptography use algorithms which
will become insecure.  This has led to a number of standards organisations
(including NIST, ETSI and ISO/IEC JTC 1/SC 27/WG 2) to start work on developing
a suite of asymmetric cryptographic algorithms which will remain secure in a
future 'post-quantum' world.

To date, the NIST work has been most influential, with the
results of its Post-Quantum Cryptography Standardization
project being closely followed by other standards bodies such
as SC 27/WG 2. The scope and status of this project are best
summarised by quoting the project web
page\footnote{\url{https://csrc.nist.gov/Projects/post-quantum-cryptography/Post-Quantum-Cryptography-Standardization}}.
The following call was announced in November 2017.
\begin{quote}
NIST has initiated a process to solicit, evaluate, and
standardize one or more quantum-resistant public-key
cryptographic algorithms. Currently, public-key cryptographic
algorithms are specified in FIPS 186-4, Digital Signature
Standard, as well as special publications SP 800-56A Revision
2, Recommendation for Pair-Wise Key Establishment Schemes Using
Discrete Logarithm Cryptography and SP 800-56B Revision 1,
Recommendation for Pair-Wise Key-Establishment Schemes Using
Integer Factorization Cryptography.  However, these algorithms
are vulnerable to attacks from large-scale quantum computers
(see NISTIR 8105, Report on Post Quantum Cryptography). It is
intended that the new public-key cryptography standards will
specify one or more additional unclassified, publicly disclosed
digital signature, public-key encryption, and key-establishment
algorithms that are available worldwide, and are capable of
protecting sensitive government information well into the
foreseeable future, including after the advent of quantum
computers.
\end{quote}

As a first step in this process, NIST solicited public comment
on draft minimum acceptability requirements, submission
requirements, and evaluation criteria for candidate algorithms.
The comments received were posted, along with a summary of the
changes made as a result of these comments. This call led to a
large number of proposals for novel algorithms, which are in
the process of being reduced to a standardisable set via a
series of rounds involving public comment and cryptanalysis.
The Round 2 candidates were announced in January 2019.

\section{5G security --- a quick summary}  \label{5G}

\subsection{Overview}  \label{overview}

In our discussion we assume that a user has established a
subscription with a network (the \emph{home network}), and that
the user's mobile phone is connecting to a \emph{serving
network}. Of course, the two networks could be the same, but in
general they are distinct.

For the purposes of this discussion, 5G security is as defined
in release 15 (R15) of the 3GPP 5G security specifications
\cite{33.501v15.6.0}.  Note that R15 is now stable, and work
has recently started on replacing it with R16; however, very
little of the system we describe looks set to be modified in
R16. 5G security as currently defined can be regarded as an
evolution of 4G security, which was itself an evolution of the
security provisions in 3G (UMTS) and 2G (GSM)\@.  However,
three major differences from 4G are:
\begin{itemize}
\item \emph{flexibility in authentication method}: the \emph{user equipment
    (UE)}, e.g.\ a mobile phone, can be authenticated to the serving
    network (formally the \emph{SEcurity Anchor Function (SEAF)} of this
    network, which cooperates with the network's \emph{AUthentication
    Server Function (AUSF)} to handle security functions) using either the
    5G AKA (Authentication and Key Agreement) protocol, itself an evolution
    of 4G AKA, or the Internet EAP-AKA$'$ protocol\footnote{EAP-AKA$'$
    (defined in RFC5448 \cite{RFC5448}) is an improved version of EAP-AKA
    (see RFC 4187, \cite{RFC4187}), and differs from EAP-AKA in that the
    keys derived as a result of its operation are bound to the name of the
    access network.};
\item \emph{robust mobile identity confidentiality}: the use of public key
    encryption removes the need to ever send the permanent user identity
    across the network in cleartext; and
\item \emph{data integrity protection}: this covers the use of keys derived
    from the session key established during authentication to protect the
    integrity of data sent across the air interface (including digitised
    voice).
\end{itemize}
We focus here on the case where the 5G AKA protocol is used, but clearly a
complete analysis would also need to consider the EAP-AKA$'$ case.  As
described below, the security provisions can be divided into four main parts:
\begin{itemize}
\item \emph{AKA}: this involves an exchange between the UE
    and the serving network, as a result of which the two
    parties are mutually authenticated and a secret session
    key is established between the two entities (see \S6.1
    of \cite{33.501v15.6.0});
\item \emph{Key derivation}: involving the derivation of
    purpose-specific keys from the session key established
    during AKA (see \S6.2 of \cite{33.501v15.6.0});
\item \emph{Mobile identity confidentiality}: this involves
    ensuring that the UE never sends the cleartext
    permanent user identity over the air interface, either
    by encrypting it or through the use of unlinkable
    temporary identifiers (see \S6.12 of
    \cite{33.501v15.6.0}); and
\item \emph{Session security}: in which individual secret
    keys, derived from the session key, are used to protect
    data sent across the air interface between the UE and
    the serving network (see \S6.4, \S6.5 and \S6.6 of
    \cite{33.501v15.6.0}).
\end{itemize}
Underlying all security services is the requirement for a subscriber to be in
possession of a card-based \emph{Universal Subscriber Identity Module (USIM)},
which can be either removable or hard-wired into the UE\@.  This USIM stores a
long-term 128-bit\footnote{See \S 5.1.7.1 of TS 33.105 \cite{33.105v15.0.0}.
Slightly surprisingly, \S 6.2.2.1 of 3GPP TS 33.501 \cite{33.501v15.6.0} allows
the key $K$ to be 128 or 256 bits long --- clearly, the intention is to permit
the change to 256-bit keys proposed in this paper, but this modification has
not been made in all the relevant specifications.} secret key $K$ for use in
AKA, which is also stored by the home network's \emph{Authentication credential
Repository and Processing Function (ARPF)}. The USIM also stores the home
network's public key used for identifier encryption (see \S\ref{IDconf} below).

Finally, note that the descriptions below are somewhat
simplified, with the goal of giving enough information for the
purposes of the subsequent analysis. In particular we have not
addressed the extensive provisions in the specifications for
security issues relating to handover and roaming.  We also do
not separate the functions of the SEAF from the associated
\emph{Access and Mobility Management Function
(AMF)}\footnote{Confusingly, AMF is also used as an
abbreviation for \emph{Authentication Management Field}, which
is another reason for not discussing the Access and Mobility
Management Function further here.}, which in any case is
required to be co-located with the SEAF (see \S 6.2.2.1 of TS
33.501 \cite{33.501v15.6.0}).

\subsection{Authentication and key agreement}

The 5G AKA protocol can be thought of as the core of 5G security.  It is very
similar to the 4G AKA protocol; one minor change is to provide the home network
with a `proof of successful authentication' of the UE to the serving network.

\subsubsection{Authentication vectors}  \label{AVs}

As in all previous system generations (2G--4G), the home network's ARPF
provides the serving network's AUSF with a 5G HE \emph{Authentication Vector
(AV)} on request. This 5G HE AV is generated in the following way (note it is
called a 5H HE AV to distinguish it from the 5G AV which is calculated from the
5G HE AV by the AUSF of the serving network; the 5G AV is what is actually used
in the protocol.

\begin{figure}[htb]
\centering
\includegraphics[width=90mm]{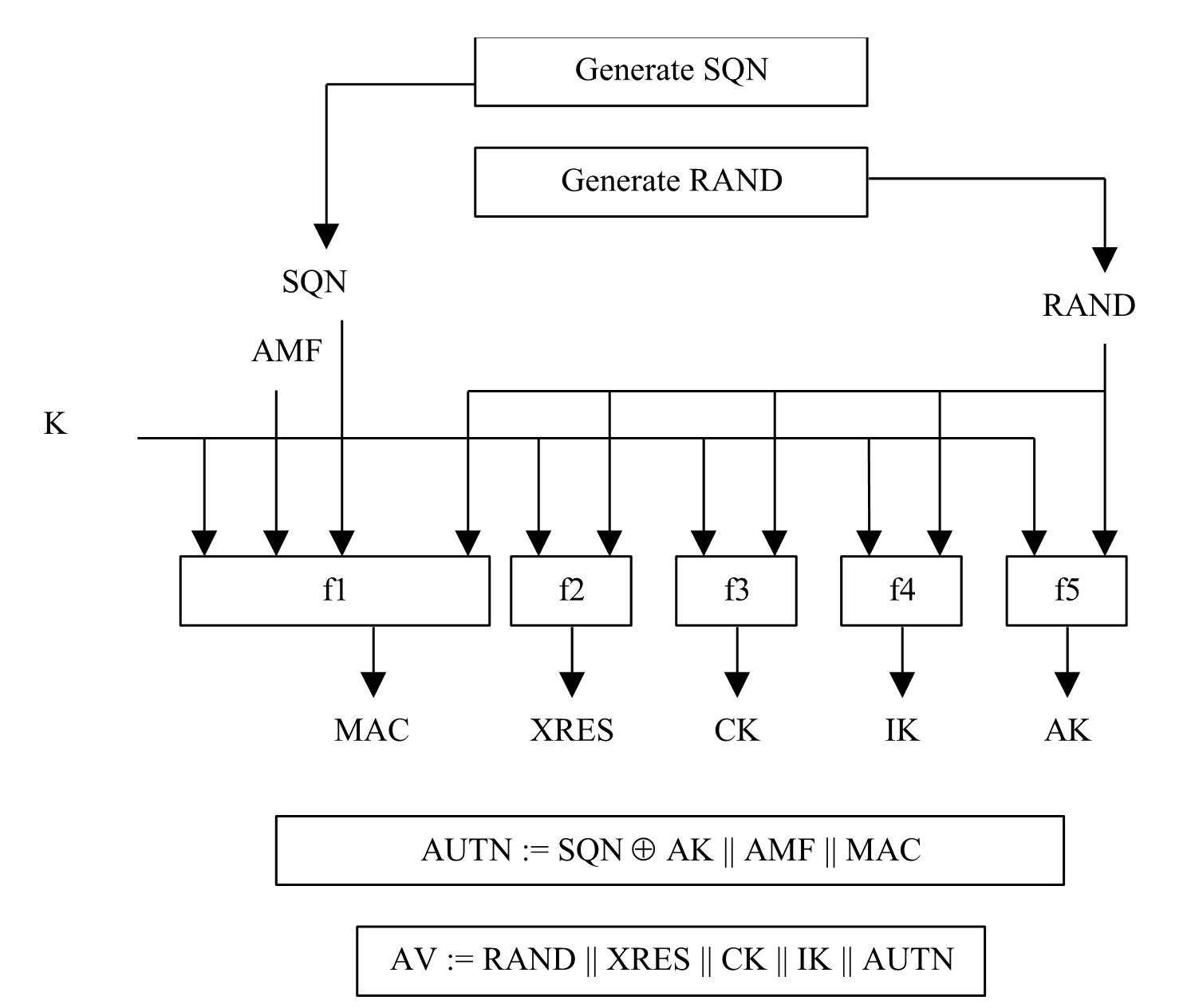}
\caption{Generation of 3G/4G AVs (Figure 7 of TS 33.102 \cite{33.102v15.1.0})}
\label{fig-AVs}
\end{figure}

A 5-value vector (RAND, XRES, CK, IK, AUTN) is first computed
in precisely the same way as AVs for 3G and 4G are generated
(see 3GPP TS 33.102 \cite{33.102v15.1.0}).  The calculation of
a 3G/4G AV is summarised in Figure~\ref{fig-AVs}. More
precisely: RAND is a 128-bit random value; XRES, CK and IK are
128-bit values generated as a function of RAND and the
subscriber secret key $K$; and AUTN is a 128-bit
\emph{authentication token}. AUTN includes a 48-bit string
SQN$\oplus$AK, where SQN is a sequence number derived from a
counter managed by the ARPF, and AK is a 48-bit encrypting
string generated as a function of RAND and the subscriber
secret key $K$.  AUTN also includes two other fields: the
16-bit Authentication Management Field (AMF) and a 64-bit
message authentication code (MAC)\@. The MAC is computed as a
function of RAND, AMF, SQN and the subscriber secret key $K$.
Of the 16 bits of the AMF, eight are available for proprietary
purposes, and could, for example, be used to indicate the suite
of algorithms $f1$--$f5$ that are in use (see below); of the
other eight, seven are reserved for future use. For further
details of use of the AMF see Annexes F and H of 3GPP TS 33.102
\cite{33.102v15.1.0}.

The functions used to derive MAC, XRES, CK, IK and AK are
denoted by $f1$, $f2$, $f3$, $f4$ and $f5$, respectively, and
are secret-key based MAC and key derivation mechanisms.  The
choice of these algorithms is left to the mobile operator,
since they are only implemented in the USIM and the ARPF, both
controlled by the operator. However, requirements for these
algorithms are defined in 3GPP TS 33.105 \cite{33.105v15.0.0}
and a default set of algorithms is provided in 3GPP TS 35.205
\cite{35.205v15.0.0} (and related documents).

Two additional computations are required to create a 5G HE AV from this vector.
\begin{itemize}
\item As described in Annex A.4 of 3GPP TS 33.501
    \cite{33.501v15.6.0}, a 128-bit value XRES* is
    generated from a combination of XRES, RAND, CK, IK and
    the serving network name, using a key derivation
    function KDF\@.  The function KDF is specified in Annex
    B.2.0 of 3GPP TS 33.220 \cite{33.220v15.4.0}, and is
    based on HMAC-SHA-256, \cite{ISO10118-3:2018,RFC2104}.
\item The values CK, IK, and the encrypted SQN from the AUTN are combined
    with an identifier for the serving network to generate a 256-bit key
    $K_{\text{AUSF}}$, using the same function KDF.
\end{itemize}
The 5G HE AV is the 4-tuple (RAND, AUTN, XRES*, $K_{\text{AUSF}}$).  On receipt
of this AV, it is used by the serving network AUSF to calculate the 5G AV in
the following way.
\begin{itemize}
\item The 128-bit value HXRES* is computed as the
    (truncated) SHA-256 hash of the concatenation of RAND
    and XRES* (see Annex A.5 of \cite{33.501v15.6.0}).
\item The 256-bit key $K_{\text{SEAF}}$ is set to the output of the
    function KDF when given as input the serving network name and
    $K_{\text{AUSF}}$ (see Annex A.6 of \cite{33.501v15.6.0}).
\end{itemize}
The 5G AV is the 4-tuple (RAND, AUTN, HXRES*, $K_{\text{SEAF}}$).  A high-level
overview of the generation of the 5G AV is provided by steps 1--4 of
Figure~\ref{fig-AKA}.

\subsubsection{The AKA protocol}  \label{AKA}

\begin{figure}[htb]
\centering
\includegraphics[width=120mm]{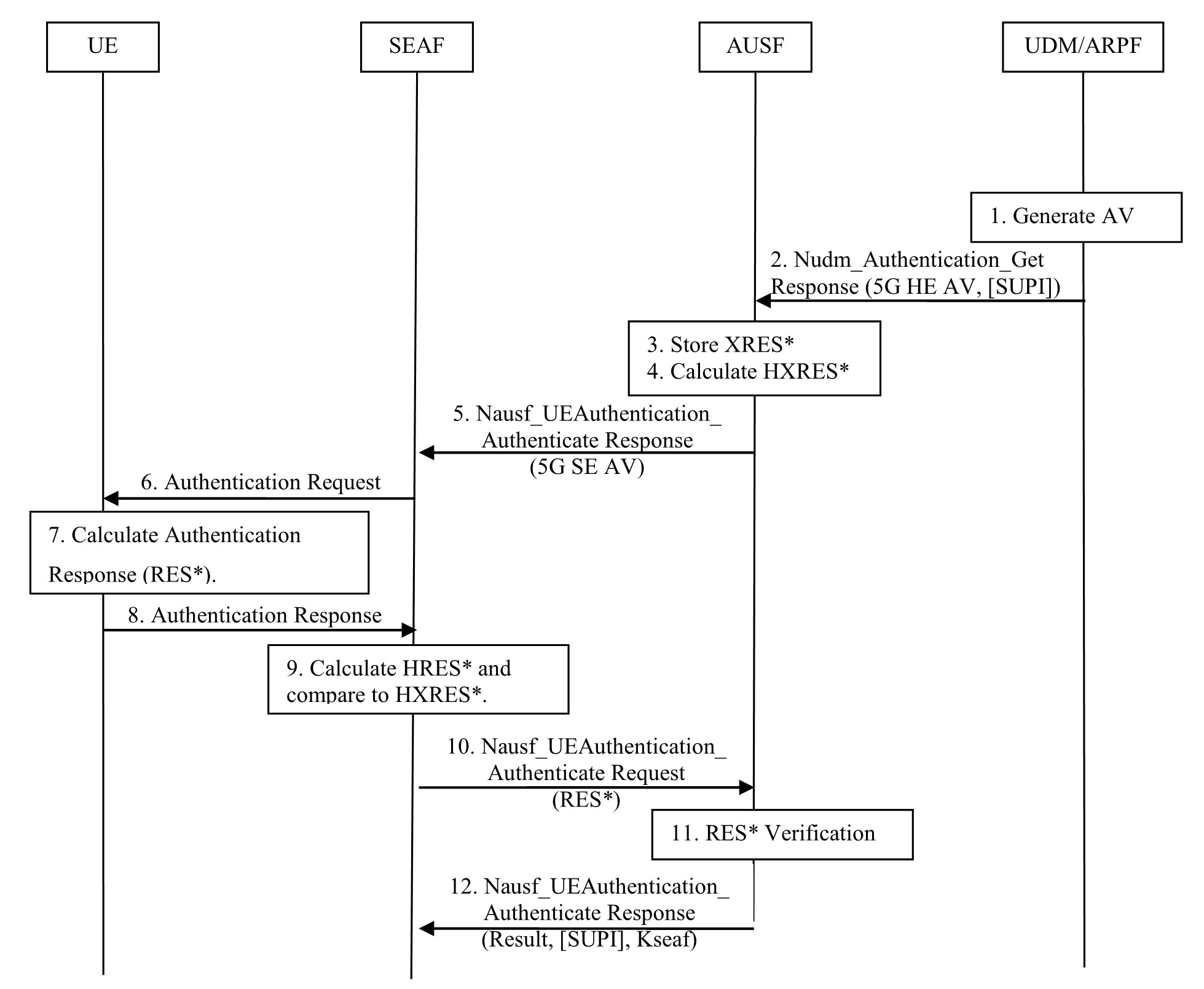}
\caption{5G AKA procedure (Figure 6.1.3.2-1 of TS 33.501 \cite{33.501v15.6.0})}
\label{fig-AKA}
\end{figure}

The AKA protocol relies on the 5G AV\@.  A summary of the 5G AV
generation process and its use in AKA is given in
Figure~\ref{fig-AKA}; the step numbers in the description below
match those used in the figure.

\begin{enumerate}
\item[5.] The first three elements of the 5G AV are passed by the AUSF to
    the SEAF.
\item[6.] The SEAF sends RAND and AUTN to the UE in an \emph{Authentication
    Request} message.  The receiving \emph{mobile equipment} (\emph{ME})
    submits the values to its resident USIM (where the ME and the USIM
    collectively make up the UE).
\item[7a.] The USIM computes AK from its secret key $K$ and
    RAND, and uses this to recover the sequence number SQN
    from AUTN\@.  SQN is checked for freshness using state
    information held by the USIM (see Annex C.2 of 3GPP TS
    33.102 \cite{33.102v15.1.0})).  The MAC in AUTN is also
    checked using the secret key $K$.
\item[7b.] If the checks succeed, the USIM calculates a response RES,
    together with keys CK and IK, using exactly the same process as used in
    computing the 3G/4G AV (see \S\ref{AVs}); these values are passed back
    to the ME, along with an indication of success or failure.
\item[7c.] The ME then computes RES* from RES, and $K_{\text{AUSF}}$ from
    CK, IK and the AUTN, using the same processes as employed to compute
    the 5G HE AV (see \S\ref{AVs}).  The ME also computes $K_{\text{SEAF}}$
    from $K_{\text{AUSF}}$ (also as described in \S\ref{AVs}).
\item[8.] The ME then returns RES* to the SEAF in an \emph{Authentication
    Response} message.
\item[9.] The SEAF computes HRES* from the received RES*
    (again as described in \S\ref{AVs}) and compares HRES*
    with HXRES*\@.  If they agree the UE is deemed
    authenticated.
\item[10--12.] The SEAF passes the receives RES* to the AUSF for further
    checks (notably to verify whether the AV is still current), and the
    AUSF passes final confirmation of the success or failure of AKA back to
    the SEAF.
\end{enumerate}

As a result of successful completion of the protocol, the SEAF
and UE are mutually authenticated and possess the authenticated
shared secret key $K_{\text{SEAF}}$.  How this is used is
described immediately below.

\subsection{Key derivation}  \label{keys}

\begin{figure}[htb]
\centering
\includegraphics[width=120mm]{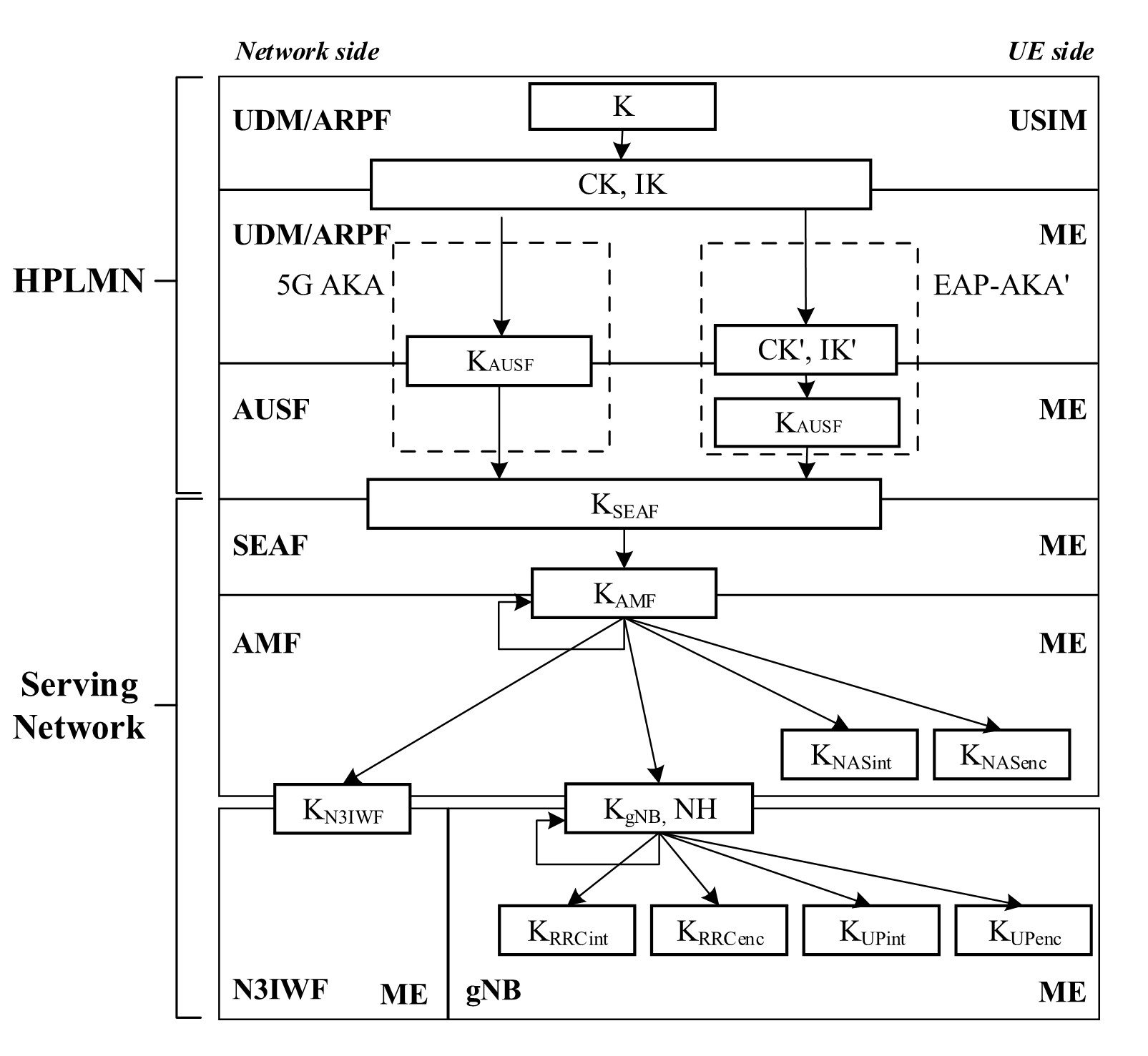}
\caption{Key derivation hierarchy (Figure 6.2.1-1 of TS 33.501 \cite{33.501v15.6.0})}
\label{fig-keys}
\end{figure}

We next provide a simplified summary of how special-purpose
keys are derived from the \emph{anchor key} $K_{\text{SEAF}}$.
We describe the SEAF (network side) key derivations; the
computations performed by the UE are identical. The key
generation hierarchy is summarised in Figure~\ref{fig-keys}.
Use of the generated keys is further described in
\S\ref{session-sec}.

Immediately following successful completion of AKA, the SEAF derives a 256-bit
key $K_{\text{AMF}}$ from the anchor key $K_{\text{SEAF}}$, and
$K_{\text{SEAF}}$ is then deleted.  To support mobility, a new key
$K'_{\text{AMF}}$ can be derived from $K_{\text{AMF}}$ for transfer to another
SEAF.

The SEAF\footnote{Actually, the AMF --- see \S\ref{overview}.} then generates a
number of operational keys, as follows.
\begin{itemize}
\item It generates a 256-bit key $K_{\text{gNB}}$ from $K_{\text{AMF}}$.
    This key is then subsequently used to generate a pair of keys
    ($K_{\text{UPint}}$, $K_{\text{UPenc}}$) for protecting the integrity
    and confidentiality, respectively, of \emph{User Plane (UP)} traffic,
    i.e.\ data and voice traffic sent between the UE and the network.  The
    key $K_{\text{gNB}}$ is also used to generate a pair of keys
    ($K_{\text{RRCint}}$, $K_{\text{RRCenc}}$) for protecting the integrity
    and confidentiality, respectively, of \emph{Radio Resource Controller
    (RRC)} signalling traffic.
\item It generates a pair of keys ($K_{\text{NASint}}$,
    $K_{\text{NASenc}}$) from $K_{\text{AMF}}$ for protecting the integrity
    and confidentiality, respectively, of \emph{Non Access Stratum (NAS)}
    layer traffic, i.e.\ NAS layer signalling traffic sent between the UE
    and the network.
\item Finally, as specified in \S 6.2.2.1 of TS 33.501
    \cite{33.501v15.6.0}, other operational keys are derived by the SEAF
    for use in different contexts; however, these are outside the scope of
    our simplified description.
\end{itemize}

The algorithms used for the key derivations described above are
specified in Annex A of 3GPP TS 33.501 \cite{33.501v15.6.0};
just as in the generation of the 5G AV (see \S\ref{AVs}), these
are all based on the key derivation function specified in Annex
B.2.0 of 3GPP TS 33.220 \cite{33.220v15.4.0}, which is based on
HMAC-SHA-256, \cite{ISO10118-3:2018,RFC2104}.

It is important to observe that all keys derived from CK and IK (from
$K_{\text{AUSF}}$ downwards in the hierarchy) are 256 bits long.  However, the
keys actually used operationally, although derived as 256-bit strings, are
truncated to 128 bits before use, since the algorithms employed only take
128-bit inputs.  The decision to use 256-bit keys within the key derivation
hierarchy dates back to 2007, \cite{S3-070733,S3a070922}, and is already built
into 4G systems.  The motivation for this feature was to enable a move to
256-bit keys at a later date.

\subsection{Mobile identity confidentiality}  \label{IDconf}

In 5G, the permanent user identifier is called the Subscription Permanent
Identifier (SUPI), and this is never sent in cleartext across the user
interface. Instead, when initiating authentication, the UE identifies itself
using either an encrypted version of the SUPI (the \emph{Subscription Concealed
Identifier (SUCI)}) or, if one is available, a previously established temporary
identifier (the \emph{5G-Globally Unique Temporary UE Identity (GUTI)}\@.
Transmission of a SUCI may also be requested by the serving network, if it
cannot resolve the 5G-GUTI.

The SUCI is created from the SUPI using a `protection scheme' (typically an
asymmetric encryption scheme) and the public key of the subscriber's home
network, where the protection scheme is either ECIES (an elliptic curve
encryption scheme (see Annex C.3 of \cite{33.501v15.6.0})) or a scheme of the
home network's choice.  The specification of ECIES follows the SECG
specifications\footnote{SECG is a US private company - see
\url{www.secginc.com}.}, rather than those of a formal standard.  The
encryption is permitted to take place in either the USIM or the ME, at the
choice of the USIM\@.  Of course, if the encryption takes place in the ME, then
a standardised scheme needs to be used. The public key of the home network is
required to be available to the UE via a trusted source, e.g.\ by being
provisioned in a USIM\@. When sending the SUCI to the serving network, the UE
also sends the identifier of the home network. Note that the encryption is
randomised, so that two SUCIs generated from the same SUPI will be distinct and
unlinkable (except, of course, by the home network which can decrypt them).

The SUCI is then used by the serving network to request authentication
information from the appropriate home network.  The home network can recover
the SUPI from the SUCI using its private decryption key. When a new 5G-GUTI is
sent to the UE across the air interface, it is sent across an encrypted link
(see \S\ref{session-sec} below).

\subsection{Session security}  \label{session-sec}

We conclude this summary of 5G security mechanisms by
considering how the various keys are used to protect data of
various types sent to and from the UE\@. As previously, this is
a somewhat simplified description, with certain special cases
ignored.

We first describe the operation of NAS data encryption and
integrity protection (see \S 6.4 of 3GPP TS 33.501
\cite{33.501v15.6.0}). Encryption and integrity protection are
applied on a per message basis. Encryption operates using a
stream cipher, and integrity protection involves the
computation of a 32-bit MAC over the \emph{unencrypted}
message.  Both encryption and MAC computation involve the use
of COUNT, a 32-bit counter value.  Both sender and receiver
maintain two counters, for traffic in the two directions.

The stream cipher keystream is computed as a function of COUNT, the cipher key
$K_{\text{NASenc}}$, a bearer identifier, and a bit indicating the direction of
transmission (uplink or downlink). Similarly, the MAC is computed as a function
of COUNT, the integrity key $K_{\text{NASint}}$, a bearer identifier, a bit
indicating the direction of transmission, and the message.

The algorithms to be used are specified in Annex B of 3GPP TS 33.401
\cite{33.401v15.9.0}.  Three possibilities are specified for the stream cipher
keystream generator, all of which take a 128-bit key: SNOW 3G (known as
128-EEA1), 128-bit AES \cite{ISO18033-3:10} in counter mode
\cite{ISO10116:2017} (128-EEA2), and ZUC (128-EEA3). Similarly, three
possibilities are specified for the MAC function, all taking a 128-bit key,
based on SNOW 3G, AES \cite{ISO18033-3:10} in CMAC mode \cite{ISO10116:2017},
and ZUC.

RRC integrity and confidentiality protection work in a very
similar way (see \S 6.5 of 3GPP TS 33.501
\cite{33.501v15.6.0}), except that the keys used are
$K_{\text{RRCint}}$ and $K_{\text{RRCenc}}$.

Analogously, integrity and confidentiality protection of UP \emph{protocol data
units (PDUs)} (see \S 6.6 of 3GPP TS 33.501 \cite{33.501v15.6.0}), works in the
same way (and using the same algorithms) except that they employ the keys
$K_{\text{UPint}}$ and $K_{\text{UPenc}}$.

\subsection{Backwards compatibility}

Before proceeding to our post-quantum analysis, we observe that the design of
the scheme, as well as being evolutionary in nature, intrinsically allows
backward compatibility in a variety of ways. In particular, the nature of the
exchange between the ME and the embedded USIM is essentially the same in 3G, 4G
and 5G. As described in step 7b of \S\ref{AKA}, the USIM passes the pair of
128-bit 3G keys CK and IK to the ME, and the ME is then responsible for
processing them appropriately to derive the session keys used for (4G and) 5G.
As we describe in \S\ref{steps}, this feature allows a way of transitioning to
a PQ-era-secure system without changing the USIM/ME interface --- this could
significantly simplify necessary future security system migrations.

\section{Post-quantum analysis}  \label{analysis}

\subsection{Keys and key derivation}  \label{analysis-keys}

As should be clear from the above descriptions, for pragmatic
backwards-compatibility reasons, the foundation of 5G security is exactly as it
was in UMTS and LTE (3G and 4G) systems.  A 128-bit key $K$, shared by the USIM
and the issuing network, is used as the foundation of all 5G security functions
(with the exception of mobile identity confidentiality, which relies on
asymmetric cryptography).  Clearly this means that if the USIM-resident key $K$
is ever discovered, then the entire basis of 5G security is undermined, despite
the fact that 256-bit keys are used in the key derivation hierarchy.

More specifically, to support the AKA protocol, $K$ is used to
help create a 5-value authentication vector, exactly as used in
UMTS, including RAND, MAC and two 128-bit keys CK and IK. These
values are then used to generate further values, as follows.
\begin{itemize}
\item The value XRES* is computed as a function of XRES,
    RAND, CK, IK and an identifier from the serving
    network.  In turn HXRES* is then derived from XRES* and
    RAND\@.  That is, HXRES* is a function solely of RAND,
    $K$ and the serving network name.
\item The keys CK and IK are then input to a further key derivation
    function (along with the encrypted SQN) to generate the 256-bit key
    $K_{\text{AUSF}}$\@.  Next, the 256-bit key $K_{\text{SEAF}}$ is
    derived from a combination of $K_{\text{AUSF}}$ and the serving network
    identifier, and the 256-bit key $K_{\text{AMF}}$ is, in turn, derived
    from $K_{\text{SEAF}}$. Finally, the 128-bit keys actually used for
    encrypting and integrity-protecting traffic are derived from
    $K_{\text{AMF}}$ (possibly as a multi-step process), initially as
    256-bit keys which are then truncated before use. It should thus be
    clear that the operational keys are a function solely of RAND, $K$, the
    serving network name (and certain other public parameters).
\end{itemize}
It should immediately be apparent that there is a major
security issue in the PQ era.  That is, security is based on a
128-bit secret key, and thus Grover's algorithm (see
\S\ref{grover}) could reduce the effective level of security to
that provided by a 64-bit key.  Searches of size $2^{64}$,
whilst not trivial, are well within the bounds of what is
possible today.  Detailed implications for the AKA protocol and
operational protection are discussed below.

It could be argued that parts of the functions used to generate the values
XRES*, HXRES* and the operational keys are issuing-network-proprietary (namely
$f1$--$f5$).  However, this gives very little additional security for two
reasons. Firstly `default' values for these functions are standardised and are,
presumably, in use in some networks.  Secondly, all such functions are
necessarily built into every USIM, and so a determined adversary could reverse
engineer the USIM to learn these functions.

Finally, and perhaps rather ironically, all current implementations will
immediately become non-compliant with the underlying standards.  As noted
above, the normative requirements for the functions $f1$--$f5$ are specified in
\S 5.1.6 of 3GPP TS 33.105 \cite{33.105v15.0.0} --- these requirements are
necessary since the functions can be chosen by the USIM issuer. For example, in
\S 5.1.6.3 it is stated that `$f3$ should be a key derivation function. In
particular, it shall be computationally infeasible to derive $K$ from knowledge
of RAND and CK'\@. However, since $f3$ takes two 128-bit inputs (RAND and $K$)
and gives the 128-bit value CK as output, Grover's algorithm will mean that in
the PQ era this requirement will be false \emph{by definition} regardless of
the choice of $f3$.

\subsection{Authentication and key establishment}

To consider in greater detail the degree of vulnerability of a 5G system in the
PQ era, suppose a malicious party has intercepted a matching pair of AKA
Authentication Request and Authentication Response messages. The first message
will contain a 128-bit RAND value, and the second will contain the 128-bit
response RES*\@. Now RES* is a function solely of RAND and $K$, and hence
Grover's algorithm means that of the order of $2^{64}$ work will be required to
deduce $K$.

That is, in the PQ era, interception of a single (successful)
AKA challenge-response pair will provide sufficient information
to determine the USIM key $K$.

\subsection{Symmetric encryption and integrity protection}

Next suppose that a malicious party has intercepted encrypted
and integrity-protected traffic sent over the network, together
with the AKA challenge RAND\@. That is, the interceptor could
obtain a set of encrypted PDUs and 32-bit MACs computed over
the corresponding decrypted PDUs. As noted above, the keystream
used for encryption is generated using a key derived from
$K_{\text{SEAF}}$, and $K_{\text{SEAF}}$ is itself a function
of the long-term key $K$.

Using this information as the basis for an attack is somewhat more difficult
than the case where a RAND and matching RES are known. The key
$K_{\text{SEAF}}$ is a function of RAND, and hence RAND would need to be
intercepted by the attacker along with the ciphertext, although not the RES
value. Known plaintext (for the encrypted PDUs) would also need to be available
to derive keystream bits, and the counter value(s) in use would also be needed.
That is, while an attack to derive $K$ is possible in theory based on analysing
encrypted data, such an attack would be far more difficult than using an
intercepted AK challenge-response pair.

It is interesting to observe that, had the MAC been computed on the encrypted
PDU as opposed to the plaintext version\footnote{The matter is not quite as
simple as sketched here.  Whilst the order of operations is `MAC then encrypt'
for user plane traffic, the NAS protocol performs `encrypt then MAC'. Both
approaches can be proven secure under certain assumptions, although `encrypt
then MAC' has certain practical and theoretical advantages making
implementation weaknesses less likely, explaining why this approach, and not
`MAC then encrypt', is standardised in ISO/IEC 19772 \cite{ISO19772C1:14}; for
further details, see, for example, Namprempre et al.'s 2014 paper
\cite{Namprempre14}.}, the MACs for intercepted PDUs could be used to conduct
an attack without any need for known plaintext bits.

\subsection{Asymmetric encryption}

We finally briefly examine the use of asymmetric encryption for
protection the secrecy of the UE permanent identity, i.e.\ the
SUPI, when sent across the radio interface.  In a typical case,
the home network's public key will be stored in the USIM and
can be extracted by any UE\@. That is, this public key is
essentially public information. Assuming use of an asymmetric
encryption technique vulnerable to the Shor algorithm (see
\S\ref{shor}), as would be the case if the `default' ECIES
scheme is employed, then discovering the private key from the
public key will be possible in the PQ era. This immediately
breaks the mobile identity confidentiality service for all
USIMs equipped with this public key --- potentially a very
large number.

\section{Future-proofing 5G security}  \label{future}

The main goal of the analysis above was to understand the
implications on the 5G security features of the advent of the
PQ era.  We now use this analysis to develop a series of
recommendations for steps that should be taken to ensure that
5G systems will be capable of offering a sufficient level of
security in the PQ era.

\subsection{Implications of the PQ era}

We start by summarising the implications of the analysis.  We
divide this analysis into two parts, namely implications for
security features relying on the USIM-stored secret key $K$
(used only with symmetric cryptography), and features relying
on the asymmetric key pairs belonging to USIM issuers.

\subsubsection{Attacks on the long-term USIM secret key}

As noted above, interception of a single AKA challenge-response
pair could be used as the basis of an attack to reveal the
secret key $K$\@.  The implications of this would include:
\begin{itemize}
\item all past (and future) intercepted encrypted voice and data traffic
    could be decrypted, as long as the RAND value sent during the
    immediately prior AKA protocol instance was also intercepted;
\item an active `man in the middle' interceptor could successfully
    manipulate data and signalling traffic by modifying the data and then
    modifying the MAC appropriately (this could, for example, lead to a
    long term denial of service if the UE configuration is modified to
    prevent use using manipulated signalling messages);
\item one or more `cloned' USIMs could be produced to
    enable mobile use at the expense of the legitimate
    account holder.
\end{itemize}

Whilst this would clearly be devastating for the security of a single
subscriber, the attack would need to be repeated for each USIM\@.  Given that
each instance of the attack would involve of the order of $2^{64}$ quantum
operations, recovering the key $K$ for more than a very small number of USIMs
might well be practically infeasible, at least in the medium term.

In addition, any sensitive data sent across the network is
likely to be encrypted at the application layer, e.g.\ using
SSL/TLS\@.  For example, use of HTTPS is becoming the norm for
all web traffic.  That is, the likelihood of the compromise of
sensitive data is probably small.  The possible future
compromise of voice traffic is, of course, a possibly sensitive
issue, but such a compromise has always been relatively easily
achieved by interfering with the land-line network.  This
possibility does not seem to have caused major concern in the
past.

Nonetheless, attacks on even a small number of `high value' individuals could
seriously damage the reputation of 5G, and there is also the possible threat
arising from large scale fraud through cloning of individual USIMs.  That is,
whilst the issue is not likely to be catastrophic for 5G, it nevertheless would
seem reasonable to see if there are simple steps that could be taken to reduce
the threat.

Finally, as noted above, there are also potential
non-conformance issues arising in the PQ era.  That is,
potentially all network implementations might be deemed
non-compliant with the standards because the functions
$f1$--$f5$ would not longer have the mandatory required
properties.

\subsubsection{Attacks on the USIM issuer key pair}

By contrast, one instance of a PQ era attack to discover a USIM issuer private
key potentially impacts the privacy of a large number of USIMs --- perhaps even
millions of subscriber accounts.  That is, the pay off in terms of effect is
large for a single instance of an attack.

However, it could be argued that this is not so serious since compromise of
such a private key would only move security back to the level that applies in
4G\@.  That is, it would allow so called \emph{IMSI catcher} attacks (see, for
example, \cite{Khan17}), where an active attacker can induce a UE to reveal the
subscriber's long-term identifier by impersonating a network. This is clearly a
privacy threat, but one that subscribers have coped with since the advent of
mobile networks.

It should also be noted that properties (arguably flaws) in the design of AKA
error messages allow a level of compromise of the subscriber identity even if
the mobile identity confidentiality feature is working.  As first pointed out
by Arapinis et al.\ \cite{Arapinis12}, AKA protocol error messages can be used
to discover whether a currently active UE is the same as a previously monitored
UE\@.  This attack works against 5G networks as well as previous generations of
mobile network systems. Essentially, when the UE responds to an Authentication
Request message, different error messages are returned if (a) the MAC in AUTN
fails the verification step, or (b) the MAC is correct but the SQN value is not
acceptable. This allows an active attacker who replays an `old' intercepted
Authentication Request message to learn whether or not the recipient of the
replay is the same as the original intended recipient of the request.

As discussed by Khan and Mitchell \cite{Khan14}, one solution
to this problem would be to merge the two error messages, i.e.\
to make the two types of error indistinguishable to an
interceptor.  However, recently Borgaonkar et al.\
\cite{Borgaonkar19} have pointed out that even this fix does
not prevent all attacks.  If any error message contains the
AUTS field, which is currently included in error messages for
case (b) above, then repeated Authentication Request replays
can be used to learn part of the SQN value for a target UE.

That is, without a careful redesign, there are issues with UE
identity privacy even when the 5G identity protection feature
is working securely.  It could thus be argued that compromise
of the USIM issuer private key, whilst clearly not desirable,
would not be a catastrophic failure.  However, there is clearly
the possibility of reputational damage, not only to the
networks, but also to all those responsible for writing and
maintaining the standards.

\subsection{Steps to be taken}  \label{steps}

Again we divide the discussion into separate cases, i.e.\ first addressing the
issues with the secret key $K$ --- in a two-phase approach --- and second
proposing ways of dealing with the threat to the USIM issuer asymmetric key
pairs.

\subsubsection{Changes to symmetric cryptography --- Phase 1}
\label{symm-changes}

It should be clear from the descriptions in \S\ref{5G} that the entire security
system is built on the use of 128-bit long-term secret keys.  Also, the
encryption and MAC functions used to protect transferred signalling and user
plane data use 128-bit keys. In an ideal world all keys and functions would be
replaced by 256-bit versions\footnote{As described above, the initial steps
towards achieving this have been in place since 4G, with the key derivation
hierarchy using 256-bit keys throughout.}. This would remove any threat of a
quantum computer based attack. However, given that any changes to 5G (as well
as 2G--4G) systems will need to be evolutionary, a staged approach to achieving
this seems appropriate, aimed at reducing the threat whilst minimising the
impact on currently deployed network infrastructures, handsets and USIMs.

The first phase, described immediately below, involves a relatively small set
of changes restricted to the USIM and ARPF; these changes alone appear to
significantly reduce the current threats. Note that these changes apply to 3G,
4G and 5G systems, a major advantage since it seems likely that it will be a
long time before 5G completely replaces earlier generation systems. The second
phase, specified in the next section, proposes evolving the use of keys in the
remainder of the system to make it completely PQ-era-secure, and builds on the
first phase changes.

\begin{enumerate}
\item At the heart of the proposed phase 1 change is to switch to 256-bit
    secret USIM keys $K$ for newly issued USIMs. Currently the relevant
    standard (\S 5.1.7.1 of TS 33.105 \cite{33.105v15.0.0}) requires $K$ to
    contain 128 bits\footnote{However, as noted above, \S 6.2.2.1 of 3GPP
    TS 33.501 \cite{33.501v15.6.0} already permits $K$ to be 128 or 256
    bits long.}. Changing this to allow it to contain either 128 or 256
    bits would be a trivial change and would have no immediate impact on
    any other parts of the system. At a certain point this could be changed
    further to \emph{recommend} use of 256-bit keys.

    Since the key $K$ never leaves the USIM or the issuing network's ARPF
(and all the functions that operate on $K$ are proprietary to the issuing
network), it would be perfectly possible for an issuing network to continue
to maintain existing USIMs using 128-bit keys, whilst all newly issued
USIMs would contain 256-bit keys.  The only changes to existing
functionality would be a small upgrade to the ARPF, and small changes to
USIM operation and the USIM personalisation process.  No changes to mobile
handsets or the network infrastructure (other than the ARPF) would be
needed.

\item To support such a change the requirements on the
    functions $f1$--$f5$, as specified in TS 33.105
    \cite{33.105v15.0.0} and TS 35.205
    \cite{35.205v15.0.0}, would need to be updated to allow
    the use of a 256-bit input key. It might also be
    expeditious to change the wording currently used to
    avoid the possibility that all existing system
    implementations would by definition become
    non-standards-compliant in the PQ era (as discussed in
    \S\ref{analysis-keys}).

\item To further support this change, examples of possible functions
    $f1$--$f5$ supporting a 256-bit input key could be specified in TS
    35.205 \cite{35.205v15.0.0}. Whilst such a design would need to be done
    with appropriate care, there is no shortage of symmetric cryptographic
    algorithms which take sufficiently long parameters to be secure in the
    PQ era, as well as well-established `generic' key derivation standards
    (see, for example, ISO/IEC 18033-3 \cite{ISO18033-3:10} and ISO/IEC
    11770-6 \cite{ISO11770-6:2016}, respectively).

\end{enumerate}

The above changes could all be implemented without changing any of the deployed
infrastructure or handsets --- all that would change would be the ARPFs and
newly issued USIMs.  Changing to use of 256-bit long-term USIM keys would also
significantly reduce the threat.  A PQ era attack could only learn an
operational key associated with a single instance of the AKA protocol.  Unless
quantum computers become very powerful and very cheap, it is hard to imagine
that performing of the order of $2^{64}$ quantum computations to learn one
operational key --- which might last for just a few hours --- will be worth the
effort at any time soon (if ever). That is, making these small changes would
appear to head off the PQ era threat for many years to come, and not least the
lifetime of currently issued USIMs.

The only other point that appears worth mentioning here regards the method used
to generate the secret USIM keys $K$.  Whilst it may not be the case for any
network, it is conceivable that some USIM issuers might have chosen to derive
all the USIM keys $K$ from a single master key.  This would avoid the need to
store (and protect) a large database of USIM keys, and instead the keys could
be regenerated as required from the master key. Certainly such an approach has
been employed in the past for generating secret keys for use in EMV-compliant
credit/debit cards.  If this is the case for any networks, and if the master
key is only 128 bits long, then this could allow a PQ era attack to learn all
the USIM keys by performing a single attack of complexity $O(2^{64})$.  Any
such networks (if they exist) should shift to use of a 256-bit master key as
soon as is practicable.

{\subsubsection{Changes to symmetric cryptography --- Phase 2}
\label{symm-changes2}

Changing the entire system to use 256-bit keys will require more far-reaching
changes to the system (covering the network infrastructure and handsets).  It
will probably make most sense to leave this to future (6G) systems, which could
be designed to be PQ-era-\@secure from the ground up.  However, the timing of
such a change will be a business and regulatory rather than a technical
decision.  The important issue is that, as described below, this can be done in
a way that builds on the phase 1 changes proposed above, without requiring any
further modifications to USIMs or the ARPF\@.  That is, backward compatibility
will not be lost, which could make migration much simpler.

At the heart of the approach is the observation that the USIM passes a pair of
128-bit keys CK and IK to the ME, both of which are derived from the long-term
secret key $K$ (using functions $f3$ and $f4$).  If we assume that $K$ has been
upgraded to a 256-bit key (as per Phase 1), and $f3$ and $f4$ are designed
appropriately, then the concatenation of CK and IK can be regarded as a 256-bit
key, i.e.\ the backwards compatibility feature that builds 5G security on CK
and IK actually allows the USIM to export a 256-bit key without any changes to
the interface specification (this has long been known --- see, for example, \S
2.3 of \cite{S3a070922}). We also observe that the 256-bit \emph{anchor key}
$K_{\text{SEAF}}$ is derived from CK, IK and the SQN in a two-stage process
(involving the 256-bit intermediate key $K_{\text{AUSF}}$)\@. That is, the
anchor key, and potentially all keys derived from it, should immediately be
PQ-era-secure.

To complete the `PQ-era upgrade', it is then a simple matter of ensuring that:
\begin{itemize}
\item all the operational keys derived from the anchor key are also 256
    bits long (and not truncated to 128 bits); and
\item the encryption and MAC functions used to protect transferred data
    employ 256-bit keys.
\end{itemize}

\subsubsection{Changes to asymmetric cryptography}  \label{asymm-changes}

As discussed above, a compromise of a USIM issuer's private key
would affect a large number of users, but the practical impact
would be limited.  Nevertheless, if only for reputational
reasons, the following changes should be implemented in due
course.  Note that these changes affect 5G only.

\begin{enumerate}

\item Annex C.3 of TS 33.501 \cite{33.501v15.6.0} currently
    permits the use of proprietary schemes for the
    protection of SUPIs.  No guidance is provided on the
    choice of such schemes. Guidance could usefully be
    added in the short term recommending use of schemes
    which offer security in the PQ era.

\item Annex C.3 of TS 33.501 \cite{33.501v15.6.0} currently
    specifies ECIES (an elliptic curve encryption scheme.
    This Annex should be updated to include the
    specification of a second PQ-era-secure asymmetric
    encryption scheme, once an appropriate candidate has
    been standardised.  Such standardised schemes should
    become available within the next couple of years, once
    the NIST process described in \S\ref{replacement} above
    has come to a conclusion.

\end{enumerate}

\section{Summary and conclusions}  \label{conclusions}

In this paper we have reviewed the operation of 5G security
features and considered in detail the ramifications of the PQ
era on the effectiveness of these features.  This has led to a
series of recommended changes to the operation of 5G security,
designed to minimise both the practical impact of the PQ era
attacks and the cost of implementing these changes.

In summary, and on the assumption that quantum computers do not
become very cheap and ubiquitous, at least in the medium term,
it would appear that a small number of relatively minor changes
will have the effect that the impact of the PQ era is minimal
in terms of security risk.

The recommended changes (most important first) are given in three separate sets
intended to be implemented in different timescales. Note that a more detailed
description of these changes and their impact is given in \S\ref{steps} above.
Note also that first set of changes apply to 3G and 4G networks, as well as 5G.

\begin{enumerate}
\item \emph{Phase 1 changes to symmetric cryptography}.  These changes
    should move forward as soon as is practicable, and can be achieved
    without impacting any deployed infrastructure or mobile terminals.

    \begin{enumerate}
    \item Modify the relevant standards to allow 256-bit (as well as
        128-bit) long-term secret keys to be stored in the USIM.  (At
        some stage, 256-bit keys should be recommended).
    \item Modify the relevant standards to update the requirements for
        functions using the long-term secret USIM key as input to allow
        use of a 256-bit key.
    \item Provide examples of recommended functions that take a 256-bit
        USIM key as input.  It is important to note that a set of
        candidate functions of this type has already been specified,
        namely the Tuak functions --- see 3GPP TS 35.231
        \cite{35.231v15.1.0}.
    \item Once the above three steps are in place, advise network
        operators to change to use of 256-bit long-term USIM keys.
    \end{enumerate}

\item \emph{Asymmetric cryptography changes}.  These changes should be
    accomplished as soon as viable alternative asymmetric encryption
    algorithms are standardised, perhaps in two-three years from the time
    of writing (late 2019).

    \begin{enumerate}
    \item Provide general guidance on the adoption of proprietary
        PQ-era-secure asymmetric encryption schemes for protecting
        permanent subscription identifiers.
    \item As and when suitable PQ-era-secure asymmetric encryption
        schemes have been standardised, at least one such scheme should
        be included in the relevant 5G standard, and its adoption by
        operators should be encouraged.
    \end{enumerate}

\item \emph{Phase 2 changes to symmetric cryptography}.  These changes
    involve modifying the ways keys are used in handsets and in mobile
    networks.  Implementing them will require changes to the operation of
    mobile terminals and mobile networks, but not to USIMs or the ARPF; it
    should be possible to specify the changes in such a way to allow
    parallel use of 128-bit and 256-bit keys and functions.

    \begin{enumerate}
    \item New symmetric cryptographic functions will need to defined
        for encryption and MAC generation.  These functions should all
        use a 256-bit key as input.  In fact work is already under way
        within 3GPP to see what, if any, new functions need to be
        defined for such a move --- see 3GPP TR 33.841
        \cite{33.841v16.1.0}.  Indeed, the content of this technical
        report to some extent overlaps with this paper.
    \item The specifications need to be modified to allow use of
        256-bit encryption and MAC functions using `untruncated'
        256-bit keys), once they have been adopted.
    \end{enumerate}

\end{enumerate}

These changes are modest in scope and appear to be eminently realisable in a
phased way.  Moreover, much has already been achieved towards completing these
changes (notably the use of 256-bit keys in the key derivation chain, the
specification of the Tuak functions, and the work in 3GPP TR 33.841
\cite{33.841v16.1.0}). Standards writers, network infrastructure and handset
manufacturers, and network operators are encouraged to complete their adoption.
The sooner moves are put in place to make the necessary changes, the smaller
the number of vulnerabilities will be if and when the PQ era dawns.

\section*{Acknowledgements}

I would like to thank everyone who has provided input to the development of
this document.  I would like to particularly thank Martin Albrecht and Karl
Norrman for their helpful advice and corrections.  Of course, all remaining
errors remain my responsibility alone.

%\bibliography{Crypto}

\end{document}